\documentclass[conference]{IEEEtran}

\usepackage{cite}
\usepackage{amsmath,amssymb,amsfonts}
\usepackage{dsfont}
\usepackage{algorithmic}
\usepackage{graphicx}
\usepackage{textcomp}
\usepackage{xcolor}
\usepackage{algorithm}

\usepackage{hyperref}
\hypersetup{
    colorlinks=true,
    linkcolor=red,
    filecolor=magenta,      
    urlcolor=black,
    citecolor=blue,
}

\begin{document}

\title{Generalised Hyperbolic State-space Models for Inference in Dynamic Systems}

\author{\IEEEauthorblockN{Yaman K{\i}ndap}
\IEEEauthorblockA{\textit{Signal Processing and Communications Lab.} \\
\textit{University of Cambridge}\\
Cambridge, UK}
\and
\IEEEauthorblockN{Simon Godsill}
\IEEEauthorblockA{\textit{Signal Processing and Communications Lab.} \\
\textit{University of Cambridge}\\
Cambridge, UK}
}

\maketitle

\begin{abstract}
In this work we study linear vector stochastic differential equation (SDE) models driven by the generalised hyperbolic (GH) L{\'e}vy process for inference in continuous-time non-Gaussian filtering problems. The GH family of stochastic processes offers a flexible framework for modelling of non-Gaussian, heavy-tailed characteristics and includes the normal inverse-Gaussian, variance-gamma and Student-t processes as special cases. We present continuous-time simulation methods for the solution of vector SDE models driven by GH processes and novel inference methodologies using a variant of sequential Markov chain Monte Carlo (MCMC). As an example a particular formulation of Langevin dynamics is studied within this framework. The model is applied to both a synthetically generated data set and a real-world financial series to demonstrate its capabilities.
\end{abstract}


\section{Introduction}

In the fields of statistical signal processing and machine learning, the problem of modelling dynamically evolving phenomena can be found in various application domains including financial markets \cite{Fama1965a}, speech recognition \cite{YuDeng2016}, health monitoring \cite{QuinnWilliamsMcIntosh2009} and more recently in generative models for images \cite{NonequilibriumThermodynamics2015}. Stochastic differential equation (SDE) models of real-world dynamic phenomena offer a powerful tool for capturing the inherent uncertainty in the estimated behaviour of such systems and provide a probabilistic framework that allows for the quantification and propagation of uncertainty. Furthermore, SDE models offer a physics-based intuitive representation of the dynamics and the associated uncertainty, which can be particularly beneficial in building interpretable machine learning systems.

Typically the driving (or forcing) function, which characterises the random fluctuations from the estimated deterministic evolution, is chosen as a Brownian motion. However, the Gaussian assumption implicit in the use of Brownian motion is often not justified in real-world applications, as many systems exhibit non-Gaussian features such as asymmetric or heavy-tailed distributions. Even with Brownian-driven paths, discretisation may require complex integrals and introduce further complexity. Instead, inference algorithms that work directly in continuous-time offer an effective alternative to discrete-time algorithms for large, irregularly sampled sequential data sets by reducing sampling costs associated with grid based strategies (\cite{SarkkaSvensson2023}, Section 4.6, \cite{Godsill_Yang_2006}).

Recent work on SDE models extends their definition to incorporate non-Gaussian L{\'e}vy processes,  allowing for a wider range of behaviours including the $\alpha$-stable and Poisson processes (\cite{GodsillRiabizKontoyiannis2019,GanAhmadGodsill2021}). This extension enhances the flexibility and applicability of SDE models, enabling them to capture more complex dynamics and uncertainties in real-world systems.

In this work, we study linear SDE models driven by a generalised hyperbolic (GH) L{\'e}vy process, which specifies a broad class of stochastic processes for varying levels of non-Gaussian, heavy-tailed and asymmetric characteristics, and include the normal inverse-Gaussian, variance-gamma and Student-t processes as special subclasses (\cite{Barndorffnielsen1977InfiniteDO,ShephardBarndorffNielsen2001}). A point process simulation algorithm for the GH process has been recently presented in \cite{kindap2023point}. Building on this algorithm, we present a simulation method for SDEs driven by GH processes and a novel continuous-time filtering algorithm which enables inference in such systems.

Specifically, the linear vector SDE model is defined as
\begin{equation}
    \label{eqn:vector_SDE}
    d \mathbf{x}(t) = \mathbf{A} \mathbf{x}(t) dt + \mathbf{L} dW(t)
\end{equation}
\noindent where $\mathbf{x}(t)$ is the $D$-dimensional state vector at time $t$, $\mathbf{A}$ and $\mathbf{L}$ are system matrices, and $W(t)$ is a background-driving L{\'e}vy process. The observations are assumed to be related to the state vector through
\begin{equation}
    \label{eqn:measurement_model}
    \mathbf{y}(t) = \mathbf{H} \mathbf{x}(t) + \varepsilon (t)
\end{equation}
\noindent where $\mathbf{H}$ is the observation matrix and $\varepsilon(t)$ is assumed to be a continuous-time white Gaussian noise with covariance $\sigma^{2}_{\varepsilon}$. 

The system of equations of the form presented in Eqs. (\ref{eqn:vector_SDE}) and (\ref{eqn:measurement_model}) define a L{\'e}vy state-space model (SSM) \cite{GodsillRiabizKontoyiannis2019}. Various well-known models can be treated as special cases of this general structure by choosing the particular forms of $\mathbf{A}$, $\mathbf{L}$ and $\mathbf{H}$. Some example models are the standard linear tracking models \cite{GodsillRiabizKontoyiannis2019}, continuous-time autoregressive (CAR), the CAR moving-average (CARMA) (\cite{Godsill_Yang_2006,Brockwell_2001}) and the Ornstein-Uhlenbeck (OU) processes \cite{Barndorff_Shephard_2001}.

As a particular case, we study a L{\'e}vy SSM formulation of Langevin dynamics defined as:
\begin{equation*}
    d\dot{x}(t) = \theta \dot{x}(t) dt + dW(t)
\end{equation*}
\noindent where $\theta < 0$ such that the evolution of the process reverts to zero in  proportion to the magnitude of $\dot{x}(t)$. Langevin dynamics find application in physics and biological sciences. The linear SDE formulation of Langevin dynamics can be expressed as:

\begin{equation}
    \begin{bmatrix} 
	dx(t) \\
	d\dot{x}(t) \\
    \end{bmatrix} = \begin{bmatrix} 
                	0 & 1 \\
                	0 & \theta \\
                    \end{bmatrix} \begin{bmatrix} 
                                	x(t) \\
                                	\dot{x}(t) \\
                                    \end{bmatrix} dt + \begin{bmatrix} 
                                	0\\
                                	1 \\
                                    \end{bmatrix} d W(t)
                                    \label{eq:statespacelangevin}
\end{equation}

\noindent where by definition $dx(t) = \dot{x}(t)dt$. Hence, a L{\'e}vy SSM allows both the state evolution $\dot{x}(t)$ and the integrated state $x(t)$ to be inferred within the same framework. 

Previous works on linear vector SDE models use various particle filtering (PF) strategies for the inference of states in Eq. (\ref{eqn:vector_SDE}) (\cite{JohnsonKindap2023,GodsillRiabizKontoyiannis2019}). Here, we develop a novel framework using a variant of Sequential Markov chain Monte Carlo (SMCMC) (\cite{Septier_et_al_2009,ClaveriaGodsill2022}) for the inference of states in a general L{\'e}vy SSM and demonstrate its applicability using Langevin dynamics driven by GH processes.

The paper is organised as follows. In Section \ref{sec:sim_gh} we review and summarise generalised shot noise methods for the simulation of GH processes. In Section \ref{sec:sim_SDE} we review fundamental results on the simulation of vector SDEs driven by GH processes. In Section \ref{sec:inference} we present our novel inference algorithm for the GH SSM and demonstrate its applicability in Section \ref{sec:applications} with both a synthetically generated data set and a British Pound - Japanese Yen foreign exchange rate data set.

\section{Simulation of generalised hyperbolic processes}
\label{sec:sim_gh}

In this section, simulation algorithms for the generalised hyperbolic process are outlined. These algorithms form the basis of both the simulation of vector SDE models and inference algorithms that are considered in this work. The general framework adopted in this work for the simulation of GH processes is the generalised shot-noise representations of L{\'e}vy processes $W(t)$ reviewed in \cite{Rosinski_2001} such that
\begin{equation}
\label{eqn:shot_noise_gen}
    W(t)=\sum_{i=1}^\infty W_i \mathds{1}_{\{V_i\leq t\}} 
\end{equation}
\noindent where $\{ V_i \in [0,T] \}_{i=1}^{\infty}$ are independent and identically distributed (i.i.d.) uniform random variables representing the arrival time of jumps, and $\{ W_i \}_{i=1}^{\infty}$ are jump sizes independent of jump times. Based on this general methodology, detailed studies on the simulation of GH processes are presented in (\cite{GodsillKindap2021,kindap2023point}). We review the main results here for completeness.

An advantageous aspect of the GH family of distributions in practical scenarios is that they can be expressed as a sum of i.i.d. random variables. Hence the GH distribution is infinitely divisible and can be the distribution of a L\'{e}vy process at time $t=1$ \cite{Barndorffnielsen1977InfiniteDO}. The GH distribution has a five parameter probability density function (pdf) defined for random variables on the real line such that (\cite{Eberlein_2001,Eberlein2003})
\begin{align}
    f_{GH}(x) = & a(\lambda, \alpha, \beta, \delta) \left( \delta^2+(x-\mu)^2 \right)^{(\lambda - \frac{1}{2})/2} \nonumber \\ &\times K_{\lambda-\frac{1}{2}} \left( \alpha \sqrt{\delta^2 + (x-\mu)^2} \right) \mathrm{exp}(\beta(x-\mu)) 
\label{eqn:GH_pdf}
\end{align}
\noindent where
\begin{equation*}
    a(\lambda, \alpha, \beta, \delta) = \frac{(\alpha^2 - \beta^2)^{\lambda/2}}{\sqrt{2\pi} \alpha^{\lambda-\frac{1}{2}} \delta^{\lambda} K_{\lambda} (\delta \sqrt{\alpha^2 - \beta^2})}
\end{equation*}
\noindent $K_{\nu}(\cdot)$ is the modified Bessel function of the second kind with index $\nu$. The parameter $\lambda \in \mathbb{R}$ characterises the tail behaviour, $\alpha > 0$ determines the shape, $0 \leq |\beta| < \alpha$ controls the skewness, $\mu \in \mathbb{R}$ is a location parameter and $\delta > 0$ is a scale parameter. 

An essential characteristic of the GH distribution, which will play a pivotal role in this study, is its connection to the generalized inverse Gaussian (GIG) distribution. Using the parameterisation $\gamma = \sqrt{\alpha^2-\beta^2}$, the three parameter probability density function $f_{GIG}(\lambda, \delta, \gamma)$ of the GIG distribution is the mixing distribution in a variance-mean mixture of Gaussians representation of the GH distribution such that \cite{Eberlein_2001}

\begin{equation}
    f_{GH}(x) = \int_{0}^{\infty} \mathcal{N} (x; \mu + \beta u, u ) f_{GIG} \left(u; \lambda, \delta, \sqrt{\alpha^2 - \beta^2} \right) du \label{mv_mixture}
\end{equation}

\noindent where $u$ is a GIG distributed random variable. 

Let $W(t)$ be a GH process on some time interval of interest $t\in[0,T]$; the characteristic function (CF) is given by (\cite{Kallenberg_2002}, Corollary 13.8), as
\begin{align*}
    E& \left[ \exp(iuW(t)) \right]\nonumber\\& = \exp \left( t \left[\int_{\mathbb{R}\setminus \{ 0 \}} (e^{iuw} -1-iw{\cal I}(|w|<1))Q(dw) \right] \right)
\end{align*}
\noindent where $Q$ is a L\'{e}vy measure under certain constraints and $Q(dw)$ is its L\'{e}vy density \cite{kindap2023point}. Here the CF is used to define a L\'{e}vy process since the pdf may not have an analytical form for all settings.

The GIG process is an instance of a more restricted class of non-negative, non-decreasing L\'{e}vy processes $G(t)$, called subordinator processes, whose 
CF is given by:
\begin{align*}
    E& \left[ \exp(iuG(t)) \right] = \exp \left( t \left[\int_{0}^\infty (e^{iux} -1)Q_{GIG}(dx) \right] \right)
\end{align*}
where $Q_{GIG}(dx)$ characterises the density of jumps for $G(t)$ such that the expected number of jumps of size $x\in[a,b]$ is $\mu_{[a,b]}=\int_{a}^b Q_{GIG}(dx)$ and the number of jumps is a Poisson random variable with mean $\mu_{[a,b]}$.

In its simplest form, the GIG L\'{e}vy density may be expressed as (\cite{Eberlein2003}, Eq. 74)
\begin{equation*}
    Q_{GIG}(x) =\frac{2 e^{-x\gamma^2/2}}{\pi^2x}\int_0^\infty\frac{e^{-\frac{z^2x}{2\delta^2}}}{z|H_{|\lambda|}(z)|^2}dz
\end{equation*}
\noindent where $H_{\nu}(z)$ is the Bessel function of the third kind, also known as the Hankel function of the first kind. Analogous to the normal mixture representation of the GH distribution, the GH L\'{e}vy measure can be expressed as:
\begin{equation}
Q(dw)=\int_{0}^\infty {\cal N}(dw;\mu+\beta x, x)Q_{GIG}(dx)\label{eq:Q_GH_mv_mixture}
\end{equation}

Both the GH and GIG processes are examples of infinite activity processes for which $\int_{0}^\infty Q(dw)\rightarrow \infty$ such that there are almost surely an infinite number of jumps in any time interval in question which renders exact simulation of these processes intractable. Furthermore, the presence of an integral involving Bessel functions in $Q_{GIG}(x)$ is another cause of intractability for the GH process.

Generalised shot-noise methods are particularly suitable for infinite activity processes since they allow the simulation of jumps in a non-increasing order, i.e. $x_i \geq x_{i+1}$. The simulation methods in \cite{kindap2023point} produce approximate sample paths by truncating the infinite number of terms in (\ref{eqn:shot_noise_gen}) to an \textit{adaptively} selected finite number of terms by bounding the residual error using concentration inequalities. In previous works on vector SDE models driven by L\'{e}vy processes the number of terms were manually tuned by hand and hence the adaptive methodology adopted in this work greatly increases the usability of such models \cite{JohnsonKindap2023}. In addition, the residual error committed by truncation is further approximated by adding an appropriately scaled Brownian motion term with drift. 

\begin{algorithm}
\caption{Generation of the jumps of a tempered stable process with L\'{e}vy density $Q_{TS}(x) = Cx^{-1-\alpha} e^{-\beta x}$ ($x\geq 0$) where $0<\alpha<1$ is the tail parameter and $\beta\geq 0$ is the tempering parameter.}
\label{alg:tempered_stable}
\begin{enumerate}
    \item Assign $N_{TS}=\emptyset$,
    \item Generate the epochs of a unit rate Poisson process, $\{\Gamma_i;\,i=1,2,3...\}$,
    \item For $i=1,2,3...$,
        \begin{itemize}
            \item Compute $x_i=\left(\frac{\alpha\Gamma_i}{C}\right)^{-1/\alpha}$,
            \item With probability $e^{-\beta x_i}$, accept $x_i$ and assign $N_{TS}=N_{TS}\cup x_i$.
        \end{itemize}
\end{enumerate}
\end{algorithm}

In order to address the problem of intractable L\'{e}vy densities, the methods in (\cite{GodsillKindap2021,kindap2023point}) rely on simulation from a tractable dominating process with L\'{e}vy measure $Q_0$ such that $dQ_0(x)/dQ_{GIG}(x) \geq 1, \,\, \forall x \in (0, \infty)$ and subsequent rejection sampling steps with rate $dQ_{GIG}(x)/dQ_0(x)$ as in (\cite{Lewis_Shedler_1979,Rosinski_2001}) to obtain the desired jump magnitudes $\{ x_i \}$ of the subordinator process. The associated jump magnitudes $\{ w_i \}$ from the GH process may be obtained through the normal mixture representation as $w_i = \mu + \beta x_i + \sigma \sqrt{x_i} u_i$ where $u_i\overset{iid}{\sim}{\cal N}(0,1)$.

The specific dominating processes for the simulation of GIG jumps studied in (\cite{GodsillKindap2021,kindap2023point}) can be expressed as either a {\em tempered stable\/} or {\em Gamma\/} process which have simple associated simulation procedures as outlined in Algorithms \ref{alg:tempered_stable} and \ref{alg:gamma_gen}.

\begin{algorithm}
\caption{Generation of the jumps of a gamma process with L\'{e}vy density $Q_{Ga}(x) = {C}{x^{-1}}e^{-\beta x}$ ($x\geq 0$) where $C>0$ is the shape parameter and $\beta>0$ is the rate parameter.}
\label{alg:gamma_gen}
\begin{enumerate}
    \item Assign $N_{Ga}=\emptyset$,
    \item Generate the epochs of a unit rate Poisson process, $\{\Gamma_i;\,i=1,2,3...\}$,
    \item For $i=1,2,3...$,
        \begin{itemize}
            \item Compute $x_i=\frac{1}{\beta \left( \exp(\Gamma_i / C) - 1 \right)}$,
            \item With probability $(1+\beta x) \exp(-\beta x_i)$, accept $x_i$ and assign $N_{Ga}=N_{Ga}\cup x_i$.
        \end{itemize}
\end{enumerate}
\end{algorithm}

The derivation of the specific form of these dominating processes differ based on the parameter values of the subordinator GIG process. Simulation algorithms for all parameter settings, including edge cases such as the Student-t process, and their derivation are presented in \cite{kindap2023point}. Here, we review the simulation algorithms for the most general parameter setting such that $\lambda \leq -0.5$ and $\delta, \gamma > 0$ as an example. The derivation and algorithmic definitions of the adaptive truncation scheme and the residual approximation procedure from \cite{kindap2023point} are omitted here. However, these algorithms  are utilised in Section \ref{sec:applications}.

In order to avoid direct calculation of the integral in $Q_{GIG}(x)$, the approach proposed in (\cite{GodsillKindap2021,kindap2023point}) is to consider a bivariate point process $Q_{GIG}(x, z)$ on $(0,\infty) \times (0,\infty)$ which has the GIG L\'{e}vy density as its marginal, i.e. $Q_{GIG}(x)=\int_0^\infty Q_{GIG}(x,z)dz$ such that
\begin{equation}
    Q_{GIG}(x, z) = \frac{2 e^{-x\gamma^2/2}}{\pi^2x}\frac{e^{-\frac{z^2x}{2\delta^2}}}{z|H_{|\lambda|}(z)|^2} \label{Q_GIG_def}
\end{equation}
\noindent Hence, the goal is to produce joint samples $\{ x_i, z_i \}$ from the point process with intensity function $Q_{GIG}(x, z)$, and retain the samples $\{x_i\}$ as samples from $Q_{GIG}(x)$. To generate these joint samples, tractable bivariate dominating processes with intensity function $Q^{0}_{GIG}(x, z)$ are designed such that thinning with probability $Q_{GIG}(x,z) / Q^0_{GIG}(x,z)$ yields samples from the desired process $Q_{GIG}$. Both the bivariate target process $Q_{GIG}(x, z)$ and the dominating processes can naturally be factorised as a marginal and a conditional point process such that $Q_{GIG}(x, z)=Q_{GIG}(x)Q_{GIG}(z|x)$ where $Q_{GIG}(z|x)$ is a probability density that may be interpreted as a {\em marking} variable and $(x,z) \in (0,\infty)\times(0,\infty)$ form a bivariate Poisson process \cite{Kingman1992}.

\begin{algorithm}
\caption{Generation of $N_1$ with marginal intensity function $Q_{N_1}(x)$.}
\label{gen_N_1}
\begin{enumerate}
    \item $N_1={\emptyset}$,
    \item Generate a gamma process $N_{Ga}^{1}$ having parameters $a_1 = \frac{z_1}{2 \pi |\lambda| (1+|\lambda|)}$ and $\beta_1=\gamma^2/2$ using Alg. \ref{alg:gamma_gen},
    \item Generate a gamma process $N_{Ga}^{2}$ having parameters $a_2 = \frac{z_1}{2 \pi (1+|\lambda|)}$ and $\beta_2= \gamma^2/2 + z_1^{2}/(2 \delta^2)$ using Alg. \ref{alg:gamma_gen},
    \item For each $x_i \in N_{Ga}^{1} \cup N_{Ga}^{2}$ accept with probability
    \[
        \frac{(2\delta^2)^{|\lambda|} \gamma(|\lambda|, (z_1^2 x_i)/(2\delta^2)) |\lambda| (1+|\lambda|)}{x_i^{|\lambda|} z_1^{2|\lambda|} (1+|\lambda| \text{exp}\left( - z_1^2 x_i / (2\delta^2) \right) ) }
    \]
    \noindent otherwise reject and delete $x_i$,
    \item For each $x_i$, simulate a $z_i$ from a truncated square-root gamma density
    \[
        \frac{\Gamma(|\lambda|)\sqrt{\text{Ga}} (z||\lambda|,x_i/(2\delta^2))}{\gamma(|\lambda|,z_1^2x_i/(2\delta^2))} {\cal I}_{0<z<z_1}
    \]
    \item With probability
    \[
        \frac{2}{\pi |H_{|\lambda|}(z_i)|^2 \left( \frac{z_i^{2|\lambda|}}{z_1^{2|\lambda|-1}} \right)}
    \]
    \noindent accept $x_i$, i.e. set $N_1=N_1\cup x_i$, otherwise discard $x_i$.
\end{enumerate}
\end{algorithm}

The bounds used in deriving the dominating processes in \cite{kindap2023point} involve a hyperparameter $z_1$ that splits the range of $z$ values into two and controls the tightness such that $0\leq z_1 \leq \left(\frac{ 2^{1-2|\lambda|}\pi}{\Gamma^2(|\lambda|)}\right)^{1/(1-2|\lambda|)}$. In fact the dominating process for the current parameter setting may be considered as a marked point process split into three independent point processes $N_{Ga}^{1}$, $N_{Ga}^{2}$ and $N_2$. While the point processes $N_{Ga}^{1}$ and $N_{Ga}^{2}$ can be considered together as the union of two independent Gamma processes with intensity function $Q_{N_1}(x)$, the point process $N_2$ can be interpreted as a modified tempered stable process with intensity function $Q_{N_2}(x)$. The associated procedures are outlined in Algorithms \ref{gen_N_1} and \ref{gen_N_2}. Finally, the set of points $N = N_1\cup N_2$ is a realisation of jump magnitudes corresponding to a GIG process having intensity function $Q_{GIG}(x)$. The associated GH process jumps may be obtained using the normal mixture representation as discussed above.

\begin{algorithm}
\caption{Generation of $N_2$ with marginal intensity function $Q_{N_2}(x)$.}
\label{gen_N_2}
\begin{enumerate}
    \item $N_2={\emptyset}$,
    \item Generate a tempered stable process $N_{TS}$ with parameters $C=\frac{\delta}{\sqrt{2\pi}}$, $\alpha=0.5$ and $\beta=\frac{z_1^2}{2\delta^2} + \frac{\gamma^2}{2}$ using Alg. \ref{alg:tempered_stable}.
    \item For each point $x_i \in N_{TS}$, accept with probability $\Gamma(0.5,z_1^2 x_i/(2\delta^2))/(\sqrt{\pi} e^{-z_1^2 x_i /(2\delta^2)})$, otherwise reject and delete $x_i$ from $N_{TS}$.
    \item For each $x_i$, simulate a $z_i$ from a truncated square-root gamma density
    \[
        \frac{\Gamma(0.5)\sqrt{\text{Ga}} (z|0.5,x_i/(2\delta^2))}{\Gamma(0.5,z_1^2x_i/(2\delta^2))} {\cal I}_{z\geq z_1}
    \]
    \item With probability 
    \[
        \frac{2}{\pi z_i|H_{|\lambda|}(z_i)|^2}
    \]
    \noindent accept $x_i$, i.e. set $N_2=N_2\cup x_i$, otherwise discard $x_i$.
\end{enumerate}
\end{algorithm}

\section{Simulation of GH SSMs}
\label{sec:sim_SDE}

In this section, we review methods of simulation for vector SDEs driven by L{\'e}vy processes. Particularly for the GH process, a conditionally Gaussian formulation of the simulation algorithm is introduced which will form the basis of our inference algorithm discussed in the following sections.

The solution to vector SDEs, as shown in Eq. (\ref{eqn:vector_SDE}), can be expressed as \cite{oksendalSDE}
\begin{equation}
    \label{eqn:vector_SDE_solution}
    \mathbf{x}(t) = e^{\mathbf{A}t} \mathbf{x}(0) + \int_{0}^{t} e^{\mathbf{A}(t-u)} \mathbf{L} dW(u)
\end{equation}
\noindent where $e^{\mathbf{A}t}$ is the matrix exponential of $\mathbf{A}t$ and $\mathbf{x}(0)$ is an initial condition. The second term in the right hand side is a stochastic integral with respect to a L{\'e}vy process $dW$. When studying the solution for specific characterisations of $\mathbf{A}$, $\mathbf{L}$ and $dW(t)$, it is usually convenient to define the stochastic term as:
\begin{equation*}
    \mathbf{I}(\mathbf{f}_t) = \int_{0}^{T} \mathbf{f}_t(u) dW(u)
\end{equation*}
\noindent where $\mathbf{f}_t(u) = e^{\mathbf{A}(t-u)} \mathbf{L} \mathds{1}_{\{u \leq t\}}$ and $T > t$ is some arbitrary boundary point. The integral $\mathbf{I}(\mathbf{f}_t)$ can only be analytically solved when $dW$ is Gaussian. Hence in general $\mathbf{I}(\mathbf{f}_t)$ has to be approximated, e.g. using direct Euler-type discretisations as studied in \cite{Godsill_Yang_2006}. In this work, we adopt a point process representation of $\mathbf{I}(\mathbf{f}_t)$ enabled by the series representation of L{\'e}vy processes in terms of its jumps times and sizes $\{ (V_i, W_i) \}$ as in (\ref{eqn:shot_noise_gen}). When $dW(t)$ is a normal variance-mean (NVM) process,  such as the GH process, the integral admits the shot-noise series representation
\begin{equation}
    \label{eqn:stoch_integral_sum}
    \mathbf{I}(\mathbf{f}_t) = \sum_{i=1}^{\infty} \mathbf{f}_t(V_i) W_i
\end{equation}
\noindent where the jump sizes $W_i$ can be further decomposed as 
\begin{equation}
    \label{eqn:jump_sizes_as_a_fnc_of_subordinator}
    W_i = \mu_W Z_i + \sigma_W \sqrt{Z_i} U_i
\end{equation}
\noindent where $\mu_W \in \mathbb{R}$ is a skewness parameter, $\sigma_W \in (0, \infty)$ is a scale parameter, $V_i \sim \mathrm{Unif}(0,T)$ are jump times, $Z_i$ are the jumps of a subordinator process and $U_i \sim \mathcal{N}(0,1)$ \cite{costa2023generalised} such that 
\begin{equation*}
     \sum_{i=1}^n\mathbf{f}_t(V_i) W_i \overset{n\rightarrow \infty}{\rightarrow} \int_{0}^{T} \mathbf{f}_t(u)dW(u)
\end{equation*}
\noindent which converges uniformly on $(0, T]$ to the original integral as $n \to \infty$. Thus the continuous-time process $\mathbf{I}(\mathbf{f}_t)$ can be represented as a countably infinite discrete sum without additional discretisation schemes as in the Gaussian case. The proof of this convergence can be found in Theorem 4 of \cite{costa2023generalised}. 

To generate random sample paths of the state vector $\mathbf{x}(t)$, the solution shown in Eq. (\ref{eqn:vector_SDE_solution}) can be incrementally simulated on each interval $(s, t]$ as

\begin{equation}
    \mathbf{x}(t) = e^{\mathbf{A}(t-s)} \mathbf{x}(s) + \sum_{i:V_i \in (s, t]} e^{\mathbf{A}(t-V_i)} \mathbf{L} W_i
    \label{eq:forwardcollapsedsum}
\end{equation}

\noindent given an initial state vector $\mathbf{x}(s)$ and jumps $\{ (V_i, W_i) : V_i \in (s, t] \}$.

Since there are an infinite number of jumps in any interval for the GH process, the summation in Eq. (\ref{eq:forwardcollapsedsum}) must be truncated to exclude small jumps of $dW(t)$ as discussed in Section \ref{sec:sim_gh}. To improve the approximation, the residual approximation schemes for the series representation of GH processes studied in \cite{kindap2023point} can be incorporated into the SDE simulation framework. The effect of residual small jumps, defined as $R^c_t = \mathbf{I}(\mathbf{f}_t) - \mathbf{I}^c(\mathbf{f}_t)$ where $c$ is the truncation level, can be included as a standard stochastic integral driven by a Brownian motion. The moments of the residual process for the GH case are derived in \cite{kindap2023point} and the validity of this residual approximation is studied in \cite{costa2023generalised}.

Notice that for NVM processes, given the set of subordinator jump sizes $\{ Z_i \}_{i=1}^{\infty}$, $\mathbf{I}(\mathbf{f}_t)$ is conditionally Gaussian which allows for efficient simulation and inference procedures to be designed. Hence, it is useful to express the integral in Eq. (\ref{eq:forwardcollapsedsum}) directly as a function of the subordinator jump times and sizes $\{ (V_i, Z_i) : V_i \in (s, t] \}$ by substituting Eq. (\ref{eqn:jump_sizes_as_a_fnc_of_subordinator}) into (\ref{eq:forwardcollapsedsum}) such that
\begin{equation*}
    \mathbf{I}(\mathbf{f}_t) | \{ (V_i, Z_i) \} \sim \mathcal{N}(\mathbf{m}, \mathbf{S})
\end{equation*}
\noindent where
\begin{equation}
    \label{eq:mforift}
    \mathbf{m} = \sum_{i:V_i \in (s, t]} \mathbf{f}_t(V_i) \mu_W Z_i
\end{equation}
\begin{equation}
    \label{eq:Sforift}
    \mathbf{S} = \sum_{i:V_i \in (s, t]} \mathbf{f}_t(V_i) \mathbf{f}_t(V_i)^{T} \sigma_W^2 Z_i
\end{equation}
The conditionally Gaussian formulation of the integral $\mathbf{I}(\mathbf{f}_t)$ leads to a fairly straightforward conditional form for the incremental simulation of the sample paths from Eq. (\ref{eq:forwardcollapsedsum}) such that
\begin{equation}
    \label{eq:sde_transition}
    p(\mathbf{x}_t | \mathbf{x}_s, \{ (V_i, Z_i) : V_i \in (s,t] \}) = \mathcal{N} (e^{\mathbf{A}(t-s)} \mathbf{x}_s + \mathbf{m}, \mathbf{S})
\end{equation}
\noindent where the notation of the state is simplified to $\mathbf{x}_t := \mathbf{x}(t)$ for brevity.

For the Langevin model shown in (\ref{eq:statespacelangevin}), the exact forms for $e^{\mathbf{A}(t-s)}$, $\mathbf{f}_t(V_i)$ and $\mathbf{f}_t(V_i) \mathbf{f}_t(V_i)^T$ are

\begin{equation*}
    e^{\mathbf{A}(t-s)} = \begin{bmatrix}1 & \frac{1}{\theta}\left( f_t(s)-1\right) \\ 0 & f_t(s)\end{bmatrix}
\end{equation*}
\begin{equation*}
    \mathbf{f}_t(V_i) = \begin{bmatrix}\frac{1}{\theta}(f_t(V_i)-1) \\ f_t(V_i)\end{bmatrix}
\end{equation*}
\begin{equation*}
     \mathbf{f}_t(V_i)\mathbf{f}_t(V_i)^T = \begin{bmatrix} \frac{1}{\theta^2}\left( f_t(V_i)-1\right)^2 & \frac{1}{\theta} f_t(V_i)\left( f_t(V_i) - 1\right) \\ & \\ \frac{1}{\theta} f_t(V_i)\left( f_t(V_i) - 1\right) & f_t(V_i)^2\end{bmatrix}
\end{equation*}

\noindent where $f_t(x) = e^{\theta(t-x)}$ for convenience. These terms are subsequently substituted into (\ref{eq:mforift}), (\ref{eq:Sforift}) and (\ref{eq:sde_transition}) to simulate incremental sample paths from the Langevin model. This procedure is summarised in Alg. \ref{alg:forward_simulation}.

\begin{algorithm}[t]
\caption{Incremental sampling from the GH SSM, $\{\mathbf{x}(t; \mu_W, \sigma_W^2, \lambda, \delta, \gamma, \theta) : s \leq t\}$, given $\mathbf{x}_s$.}
\label{alg:forward_simulation}
\begin{enumerate}
    \item Generate jump times and sizes $\{ (V_i, Z_i) : V_i \in (s, t] \}$ from a GIG process using Algs. \ref{gen_N_1} and \ref{gen_N_2}, 
    \item Compute the moments $\mathbf{m}$ and $\mathbf{S}$ of the associated conditional stochastic integral using the jumps $\{ (V_i, Z_i) : V_i \in (s, t] \}$ and Eqs. (\ref{eq:mforift}), (\ref{eq:Sforift}),
    \item Using $\mathbf{x}_s$, sample a conditionally Gaussian random variable $\mathbf{x}_t \sim \mathcal{N} (e^{\mathbf{A}(t-s)} \mathbf{x}_s + \mathbf{m}, \mathbf{S})$.
\end{enumerate}
\end{algorithm}

\section{Inference for GH SSMs}
\label{sec:inference}

In this section we describe a novel sequential inference algorithm for the states of a L{\'e}vy SSM driven by a normal variance-mean process such as the generalised hyperbolic case. We start by briefly reviewing the recursive state estimation problem in the context of L{\'e}vy SSMs and subsequently introduce our SMCMC algorithm for this task.

The problem of filtering in a dynamic system can be described as the computation of the posterior distribution of a vector of state variables $\mathbf{x}_t$ conditional on a batch of observations $y_{0:t} := \{ y(s_i) : 0 \leq s_i \leq t \}$ \cite{CappeGodsillMoulines2007}, where here $\{s_i\}$ is a set of discrete observation times for the process. The sequential nature of dynamic systems allow the filtering problem to be solved incrementally for all $t$.

Given an initial filtering estimate at time $s$, denoted as $p(\mathbf{x}_s | y_{0:s})$, the filtering estimate at time $t>s$ can be obtained using a two step recursive algorithm involving a prediction and a correction step. As discussed later in this section, we will devise a sequential algorithm based on an approximate collapsing of $p(\mathbf{x}_s | y_{0:s})$ onto a single Gaussian, see Section \ref{sec:sim_SDE}.

Since the associated densities of a L{\'e}vy SSM involve a set of jump times and magnitudes, define $\{ (V_i, Z_i) \}_{(s,t]} := \{ (V_i, Z_i) : V_i \in (s, t] \}$ for convenience. Then, the predictive density $p(\mathbf{x}_t | y_{0:s}, \{ (V_i, Z_i) \}_{(s,t]})$ can be expressed as

\begin{equation}
    \label{eqn:prediction_step}
    p(\mathbf{x}_t | y_{0:s}, \{ (V_i, Z_i) \}_{(s,t]}) = \int f(\mathbf{x}_t | \mathbf{x}_s) p(\mathbf{x}_s | y_{0:s}) d\mathbf{x}_s
\end{equation}

\noindent where the state transition density $f(\mathbf{x}_t | \mathbf{x}_s)$ can be derived from the forward simulation of the SDE, as specified in Eq. (\ref{eq:sde_transition}). Note that the generic notation for the state transition density $f(\mathbf{x}_t | \mathbf{x}_s)$ omits the dependence on the jumps $\{ (V_i, Z_i) \}_{(s,t]}$ for notational convenience.

After an observation at time $t$ is measured, the predictive density can be updated as

\begin{equation}
    \label{eqn:correction_step}
    p(\mathbf{x}_t | y_{0:t}, \{ (V_i, Z_i) \}_{(s,t]}) = \frac{p(y_t | \mathbf{x}_t) p(\mathbf{x}_t | y_{0:s}, \{ (V_i, Z_i) \}_{(s,t]})}{p(y_t|y_{0:s}, \{ (V_i, Z_i) \}_{(s,t]})}
\end{equation}

Since the jump times and sizes $\{ (V_i, Z_i) \}_{(s,t]}$ are latent variables in a L{\'e}vy SSM, we want to marginalise over them in practice to obtain the filtering estimate at time $t$ as

\begin{align}
    \label{eqn:marginalisation_over_jumps}
    p(\mathbf{x}_t | y_{0:t}) = \int & p(\mathbf{x}_t | y_{0:t}, \{ (V_i, Z_i) \}_{(s,t]}) \nonumber \\ & \quad \quad p(\{ (V_i, Z_i) \}_{(s,t]} | y_{0:t}) d\{ (V_i, Z_i) \}_{(s,t]}
\end{align}

\noindent where $\int \, \, d\{ (V_i, Z_i) \}_{(s,t]}$ is used to indicate a marginalisation over all possible jump sequences in $(s,t]$. 

The Gaussian noise assumption for the observation density $p(y_t | \mathbf{x}_t)$ and the conditionally Gaussian formulation of the transition density allows both Eqs. (\ref{eqn:prediction_step}) and (\ref{eqn:correction_step}) to be analytically solved and the resulting densities are still conditionally Gaussian. Hence, given a set of jumps $\{ (V_i, Z_i) \}_{(s,t]}$ the filtering problem can be solved by standard Kalman filtering recursions.

Assuming that the previous filtering estimate is approximated as a single Gaussian, $p(\mathbf{x}_s | y_{0:s}) = \mathcal{N} (\mathbf{x}_s; \boldsymbol{\mu}_{s}, \mathbf{C}_{s})$, and given $\{ (V_i, Z_i) \}_{(s,t]}$, the prediction equations from time $s$ to $t$ result in a Gaussian density $\mathcal{N}(\mathbf{x}_t; \boldsymbol{\mu}_{t|s}, \mathbf{C}_{t|s})$ such that (\cite{GodsillRiabizKontoyiannis2019,JohnsonKindap2023}), 
\begin{align}
    \boldsymbol{\mu}_{t|s} &= \mathbf{F} \boldsymbol{\mu}_{s} + \mathbf{m} \label{Kalman1}\\
    \mathbf{C}_{t|s} &= \mathbf{F} \mathbf{C}_{s} \mathbf{F}^{T} + \mathbf{S}
\end{align}
\noindent where $\mathbf{F} = e^{\mathbf{A}(t-s)}$ and $\mathbf{m}$, $\mathbf{S}$ are calculated using Eqs. (\ref{eq:mforift}) and (\ref{eq:Sforift}). Then, given the observation matrix $\mathbf{H}$ and noise covariance $\sigma^{2}_{\varepsilon}$ in Eq. (\ref{eqn:measurement_model}), the Kalman correction step results in another Gaussian density $\mathcal{N}(\mathbf{x}_t; \boldsymbol{\mu}_{t|t}, \mathbf{C}_{t|t})$ such that
\begin{align}
    \mathbf{K}_t &= \mathbf{C}_{t|s} \mathbf{H}^T (\mathbf{H} \mathbf{C}_{t|s} \mathbf{H}^T + \sigma^{2}_{\varepsilon})^{-1} \label{eqn:kalman_gain}\\
    \boldsymbol{\mu}_{t|t} &= \boldsymbol{\mu}_{t|s} + \mathbf{K}_t (y_t - \mathbf{H} \boldsymbol{\mu}_{t|s} ) \label{eq:updatea}\\
    \mathbf{C}_{t|t} &= \mathbf{C}_{t|s} - \mathbf{K}_t \mathbf{H} \mathbf{C}_{t|s} \label{eq:updateC}    
\end{align}
\noindent where $\mathbf{K}_t$ is the Kalman gain. Additionally, the marginal-conditional likelihood $p(y_t | y_{0:s}, \{ (V_i, Z_i) \}_{(s,t]}, \sigma_{\varepsilon})$ can be readily evaluated during the recursions as

\begin{equation}
    p(y_t | y_{0:s}, \{ (V_i, Z_i) \}_{(s,t]}, \sigma_{\varepsilon}) = \mathcal{N}( y_t; \mathbf{H} \boldsymbol{\mu}_{t|s}, \mathbf{H} \mathbf{C}_{t|s} \mathbf{H}^{T} + \sigma^{2}_{\varepsilon}) \label{eq:marginal_conditional_likelihoood}
\end{equation}

\noindent which will be used in later stages of our inference algorithm.

While it is straightforward to solve the filtering problem given a set of jumps $\{ (V_i, Z_i) \}_{(s,t]}$ as described above, sampling from the posterior density over the latent jumps in (\ref{eqn:marginalisation_over_jumps}) is intractable and hence the estimation of the filtering density also becomes intractable. In previous works, this problem was overcome using particle filtering strategies where each simulated set of jumps $\{ (V_i, Z_i) \}_{(s,t]}^{(j)}$ were associated with an importance weight $w_j$ such that (\ref{eqn:marginalisation_over_jumps}) can be evaluated as a weighted sum over conditionally Gaussian densities $p(\mathbf{x}_t | y_{0:t}, \{ (V_i, Z_i) \}_{(s,t]}^{(j)})$. In this work, we propose a new inference framework based on Monte Carlo sampling strategies where samples from the posterior density over the latent jumps are obtained using an MCMC algorithm at each iteration.

The posterior density over the latent jumps in $(s,t]$ can be expressed through Bayes' theorem as
\begin{equation}
    p(\{ (V_i, Z_i) \} | y_{0:t}) = \frac{p(y_t | y_{0:s}, \{ (V_i, Z_i) \}) p(\{ (V_i, Z_i) \})}{p(y_t | y_{0:s})}
\end{equation}
\noindent where the likelihood term is evaluated as part of the Kalman filtering recursions and shown in (\ref{eq:marginal_conditional_likelihoood}). The prior density $p(\{ (V_i, Z_i) \}_{(s,t]})$ for the jump sequence is intractable to compute, especially for infinite activity processes such as these. Nevertheless, under the truncated model with level $c$ in Eq. (\ref{eqn:shot_noise_gen}), it would be possible to characterise this prior fully and hence use it to perform MCMC by proposing changes to individual jumps, for example. Here we leave such a possibility for future investigations and rely on forward simulation of the jumps from their prior only in the MCMC. 

A simple and effective Metropolis–Hastings (MH) algorithm targeting the posterior density over the latent jumps may be constructed
by proposing the jumps from their GIG prior $p(\{ (V_i, Z_i) \}_{(s,t]})$ using Algs. \ref{gen_N_1} and \ref{gen_N_2}. In this case MH acceptance probabilities are obtained solely in terms of 
 the marginal-conditional likelihood term $p(y_t | y_{0:s}, \{ (V_i, Z_i) \}_{(s,t]})$. For each proposed change $\{ (V_i, Z_i) \}_{(s,t]}^{(\prime)})\sim p(\{ (V_i, Z_i) \}_{(s,t]})$ to the current state of the jump sequence at iteration $j$, $\{ (V_i, Z_i) \}_{(s,t]}^{(j)}$, the Kalman filtering recursions (\ref{Kalman1})-(\ref{eq:marginal_conditional_likelihoood}) are carried out and  the new jump sequence $\{ (V_i, Z_i) \}_{(s,t]}^{(\prime)}$ is accepted with  probability
\begin{equation}
    \label{eqn:acceptance_prob}
    \alpha = \text{min} \left( 1, \frac{p(y_t | y_{0:s}, \{ (V_i, Z_i) \}_{(s,t]}^{(\prime)})}{p(y_t | y_{0:s}, \{ (V_i, Z_i) \}_{(s,t]}^{(j)})} \right)\,
\end{equation}
\noindent otherwise the move is rejected and the jump sequence remains unchanged. Such an MH algorithm is shown to be effective for producing jump samples from the posterior density in Section \ref{sec:applications}.

The chain is run to convergence, and 
the chain $\{ (V_i, Z_i) \}_{(s,t]}^{(j)}$ and their associated filtering estimates $\mathcal{N}(\mathbf{x}_t; \boldsymbol{\mu}_{t|t}^{(j)}, \mathbf{C}_{t|t}^{(j)})$ are combined to make a Gaussian mixture model approximation,
\begin{equation}
    \label{eqn:gaussian_mixture}
    p(\mathbf{x}_t|y_{0:t}) \approx \frac{1}{N} \sum_{j=1}^{N} \mathcal{N}(\mathbf{x}_t; \boldsymbol{\mu}_{t|t}^{(j)}, \mathbf{C}_{t|t}^{(j)})
\end{equation}
\noindent where $N$ is total number of iterations of the chain. This approximation is then collapsed onto a single Gaussian, in  a spirit similar to \cite{KotechaDjuric2003} for the PF, as $p(\mathbf{x}_t|y_{0:t}) \approx \mathcal{N}(\mathbf{x}_t; \boldsymbol{\mu}_t, \mathbf{C}_t)$ by moment matching,
\begin{align}
    \boldsymbol{\mu}_t &= \frac{1}{N} \sum_{j=1}^{N} \boldsymbol{\mu}_{t|t}^{(j)} \\
    \mathbf{C}_t &= \frac{1}{N} \sum_{j=1}^{N} \left[ \mathbf{C}_{t|t}^{(j)} + (\boldsymbol{\mu}_{t|t}^{(j)} - \boldsymbol{\mu}_t) (\boldsymbol{\mu}_{t|t}^{(j)} - \boldsymbol{\mu}_t)^T \right]
\end{align}
Hence Eq. (\ref{eqn:gaussian_mixture}) forms an approximation of the filtering density in Eq. (\ref{eqn:marginalisation_over_jumps}). Furthermore the posterior density over the GIG jumps may also be estimated by storing the underlying samples $\{ (V_i, Z_i) \}_{(s,t]}^{(j)}$ from the MCMC algorithm. Note that the conditionally Gaussian formulation of the stochastic integral $\mathbf{I}(\mathbf{f}_t)$ greatly simplifies the design of the filtering algorithm by enabling the use of standard Kalman filtering recursions and also reduces the variance in our estimates compared to algorithms involving direct sampling of the GH process. Furthermore, the Gaussian approximation of the mixture filtering density at $t^{\prime} > t$ in (\ref{eqn:gaussian_mixture}) allows the estimation procedure for $(t, t^{\prime}]$ to be independent of the previous jumps $\{ (V_i, Z_i) \}_{(0,t]}$, a simplification compared to more complex SMCMC and PF approaches to this problem. Note that our algorithm, involving the approximation of the filtering density by a single Gaussian, is distinct from the more familiar SMCMC approaches in the literature (\cite{Septier_et_al_2009,ClaveriaGodsill2022}), and future work will compare the performance of these algorithms and with regular particle filters \cite{JohnsonKindap2023}.

\section{Applications}
\label{sec:applications}

In this section we demonstrate the applicability of the proposed inference algorithm for L{\'e}vy SSMs driven by a GH processes. Firstly, we present results of a synthetically generated Langevin dynamics data set with known and fixed parameters. Here, we display the performance of the proposed sequential MCMC methodology in Section \ref{sec:inference} on the estimation of the state vector $\mathbf{x}(t)$. Furthermore, we apply our formulation of the Langevin dynamics to a British pound (GBP) to Japanese Yen (JPY) foreign exchange rate data set sampled on 07/01/2013.

\begin{figure}[h!]
  \centering
  \includegraphics[width=0.5\textwidth]{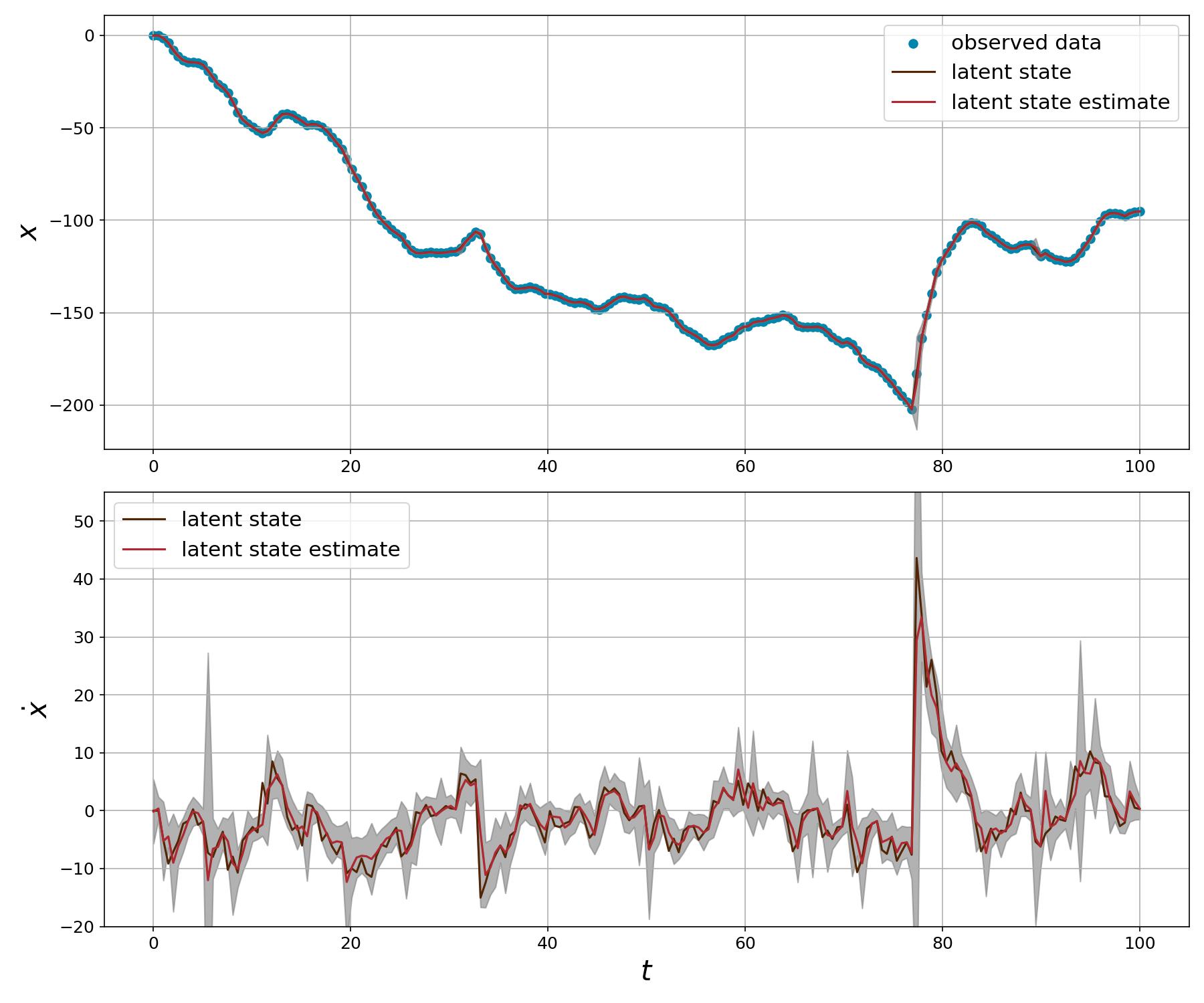}
  \caption{Estimation of states using sequential MCMC for synthetically generated data. With L{\'e}vy SSM parameters $\theta=-0.5$, $\mu=0$, $\sigma=1$, $\sigma_{\varepsilon}=0.1$ and GIG parameters $\lambda=-0.8$, $\gamma=0.01$, $\delta=1.$.}
  \label{fig:example1}
\end{figure}

For the synthetically generated data set, the parameters of the driving GH process are chosen as $\lambda=-0.8$, $\gamma=0.01$ and $\delta=1.0$ which are defined in terms of the underlying GIG subordinator process. Furthermore, $\beta=\mu_W=0$ which implies that the process is symmetric and the additional scale parameter $\sigma_W$ in Eq. (\ref{eqn:jump_sizes_as_a_fnc_of_subordinator}) is chosen to be 1.0. An independent GIG process, and its extension to a GH process, can be simulated using Algs. \ref{gen_N_1} and \ref{gen_N_2} for this parameter setting. The damping parameter $\theta$ for the evolution of the acceleration of a particle moving under Langevin dynamics is set to $-0.5$ and the observation noise variance $\sigma_{\varepsilon}^2$ is set to $0.1$. The associated SDE is initialised as $\mathbf{x}(0)=\mathbf{0}$ and the process is simulated between $(0, 100]$ with $200$ linearly spaced observations drawn according to Eq. (\ref{eqn:measurement_model}) with $\mathbf{H} = [1 \quad 0]$.

The latent states $x(t)$, $\dot{x}(t)$ and the observed data $y(t)$ are visualised in Fig. \ref{fig:example1} together with the results of our sequential MCMC filter. Notice that all of the observations are shown in the first plot with blue dots and the second plot showing the velocity of the particle is completely unobserved. Both plots include sequentially computed latent state estimates in addition to 3 standard deviation ranges in grey showing the uncertainty in our estimates. At each time interval $100$ samples are generated from the jump proposal and the associated conditional densities are accepted with probability as in (\ref{eqn:acceptance_prob}).

The results suggest that our proposed algorithm is able to accurately infer the true latent states. Specifically, the jump causing a large change in velocity around $t=78$ is successfully estimated and the model is able to track the state without any lags. The effect of this jump on $x(t)$ can also be seen as a rapid upwards change in position in the first plot. Furthermore, due to the damping effect in Langevin dynamics the velocity $\dot{x}(t)$ gradually reverts to its zero mean levels.

\begin{figure}[h!]
  \centering
  \includegraphics[width=0.5\textwidth]{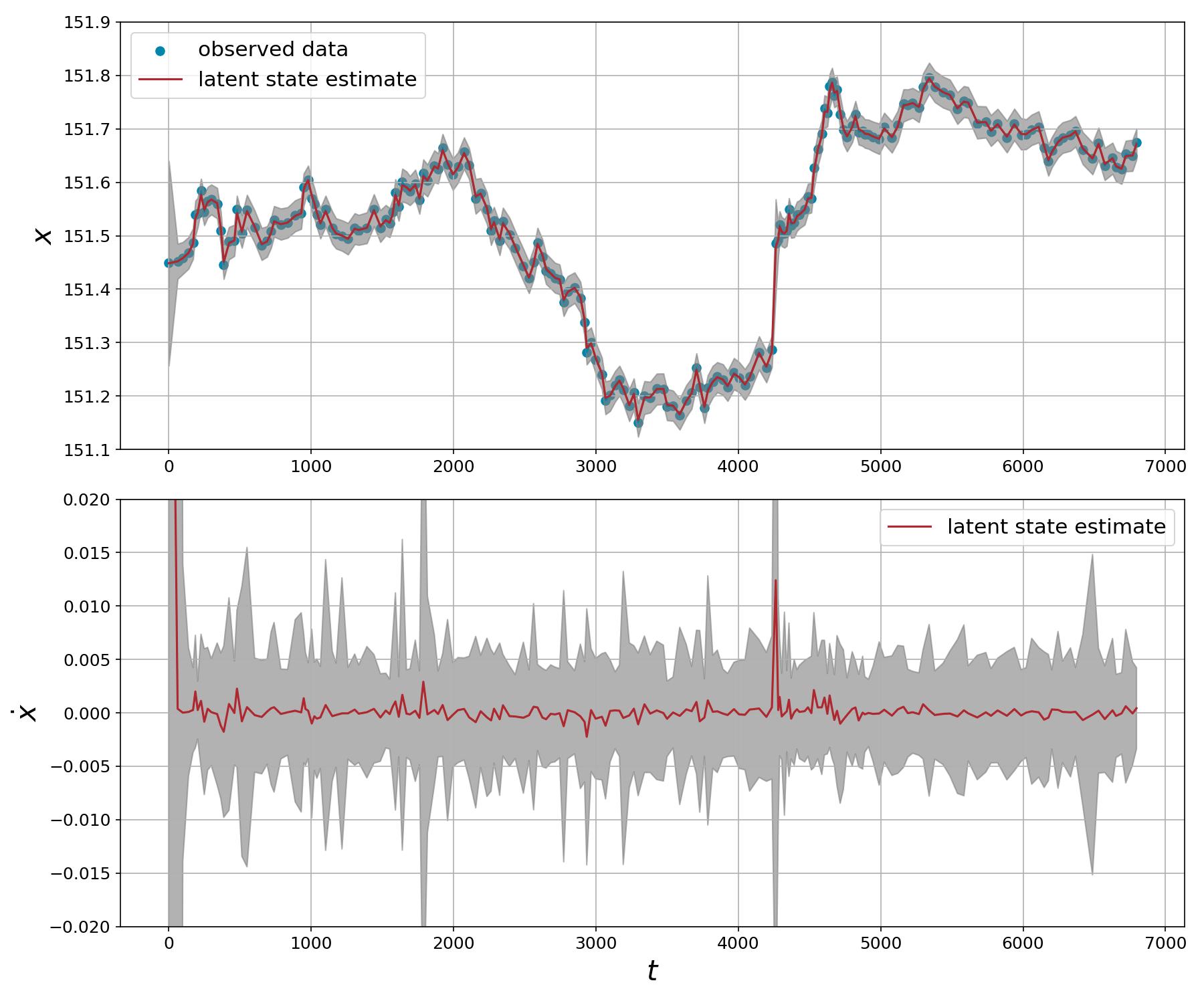}
  \caption{Estimation of states using sequential MCMC for historical intraday foreign exchange rates of British pound (GBP) to Japanese Yen (JPY) sampled on 07/01/2013. With L{\'e}vy SSM parameters $\theta=-0.1$, $\mu=0$, $\sigma=10^{-4}$, $\sigma_{\varepsilon}=10^{-2}$ and GIG parameters $\lambda=-0.6$, $\gamma=0.001$, $\delta=1$.}
  \label{fig:example3}
\end{figure}

For the foreign exchange data set, the original data was downsampled to include every 500th irregularly arriving order price for a period of $113$ minutes. The observed data used for the experiment includes $200$ points in total. The results are shown in Fig. \ref{fig:example3} where the x-axis is in seconds. The parameters of the driving GH process as well as the L{\'e}vy SSM parameters are tuned by hand using grid-based strategies that maximise the average log marginal likelihood of the data. 

The inferred velocity of the rates (lower plot) include rapid shifts, most notably around $t=4200$, which suggest that a Brownian-driven model would be inappropriate for such a heavy-tailed process. The state on the top plot is able to follow the price accurately without losing track. Hence, the ability to sequentially estimate large deviations in state will likely be of significant assistance in momentum-based trading strategies.

\section{Conclusions}

In this work, we introduced a L{\'e}vy SSM driven by a GH process which provides a flexible representation of continuous-time stochastic linear systems with non-Gaussian properties. We introduced a novel sequential inference algorithm for this model that may be readily generalised to L{\'e}vy SSM driven by normal variance-mean processes. A specific formulation of Langevin dynamics is used to demonstrate the applicability of L{\'e}vy SSMs and our sequential MCMC algorithm in Section \ref{sec:applications}. However, it is worth noting that the class of L{\'e}vy SSMs is able to represent a wider range of models such as the standard object tracking including the random walk, constant velocity, constant acceleration models. Together with the broad class of processes represented by a GH process, our work enables modelling of a wide range of real-world phenomena with applications in a variety of fields.

Future research directions on these models include improved inference algorithms that incorporate proposal distributions for individual jumps, estimation of parameters by extending the state to include time-dependent unknown parameter values and non-parametric estimation of the underlying jump process.

\bibliographystyle{IEEEtran}
\bibliography{bibliography}

\end{document}